\documentclass[letterpaper, 10 pt, conference]{ieeeconf}
\usepackage[utf8]{inputenc}
\usepackage[english]{babel}
\usepackage{amsmath,amssymb,amsfonts}
\usepackage{parskip}
\usepackage{graphicx}
\usepackage{hyperref}
\usepackage{float}
\usepackage{algpseudocode}
\usepackage{algorithm}
\usepackage {optidef}
\usepackage{array}
\usepackage{setspace}
\usepackage{cuted}
\usepackage{titlesec}
\titlespacing*{\section} {0pt}{0.5ex}{0.5ex}
\titlespacing*{\subsection} {0pt}{0.3ex}{0.3ex}
\let\labelindent\relax
\usepackage{enumitem}
\setlist{nosep}

\usepackage{xcolor}
 \definecolor{nblue}{rgb}{0,0.263,0.576}

\newcommand{\nd}[1]{\mathsf{#1}}

\newtheorem{thm}{Theorem}[section]

\newtheorem{prop}[thm]{Proposition}

\let\OLDthebibliography\thebibliography
\renewcommand\thebibliography[1]{
  \OLDthebibliography{#1}
 \setlength{\parskip}{0pt}
 \setlength{\itemsep}{0pt plus 0.3ex}
 }

\IEEEoverridecommandlockouts                      
\title{Optimal gait design for nonlinear soft robotic crawlers}
\author{Yenan Shen$^1$, Naomi Ehrich Leonard$^1$,  Bassam Bamieh$^2$, Juncal Arbelaiz$^3$
\date{\today}
\thanks{$^{1}$ YS \& NEL are with Department of Mechanical and Aerospace Engineering, Princeton University, Princeton, NJ 08540, USA
        {\tt\small yenan.shen@princeton.edu, naomi@princeton.edu }}%
\thanks{$^{2}$ BB is with the Department of Mechanical and Environmental Engineering, University of California Santa Barbara, CA 91106, USA
        {\tt\small bamieh@ucsb.edu}}
\thanks{$^{3}$ JA is with the Center for Statistics and Machine Learning and with the Princeton Neuroscience Institute, Princeton University, Princeton, NJ 08540, USA    {\tt\small arbelaiz@princeton.edu}}}

\begin{document}

\pagestyle{empty}
\maketitle
\thispagestyle{empty}

\begin{abstract}
Soft robots offer a frontier in robotics with enormous potential for safe human-robot interaction and agility in uncertain environments.  A stepping stone towards unlocking their potential is a control theory tailored to soft robotics, including a principled framework for gait design. We analyze the problem of optimal gait design for a soft crawling body  -- \textit{the crawler}. The crawler is an elastic body with the control signal defined as actuation forces between segments of the body. We consider the simplest such crawler: a two-segmented body with a passive mechanical connection modeling the viscoelastic body dynamics and a symmetric control force modeling actuation between the two body segments. The model accounts for the nonlinear asymmetric friction with the ground, which together with the symmetric actuation forces enable the crawler’s locomotion. Using a describing-function analysis, we show that when the body is forced sinusoidally, the optimal actuator contraction frequency corresponds to the body’s natural frequency when operating with only passive dynamics.  We then use the framework of Optimal Periodic Control (OPC) to design optimal force cycles of arbitrary waveform and the corresponding crawling gaits. We provide a hill-climbing algorithm to solve the OPC problem numerically. Our proposed methods and results inform the design of optimal forcing and gaits for more complex and multi-segmented crawling soft bodies. 
\end{abstract}

\section{Introduction}
\label{sec:Introduction}
The control of rhythms across scales is 
of fundamental importance for proper biological function \cite{Juarez-Alvarez2024,Sepulchre2019}.  
Rhythmic movements including walking, swimming, and crawling, span a large portion of the animal locomotive repertoire \cite{Simoni2007}. The creation of such rhythms -- typically attributed to Central Pattern Generators (CPGs) \cite{Ijspeert2008} -- requires coordinated control signals mediating the interactions of the musculoskeletal and nervous systems with the environment \cite{Chiel1997}.
Analogously, in engineering, the generation of rhythmic control signals is of utmost importance in a plethora of applications. 
In this work, we investigate one of them: rhythm design for the locomotion of soft robots. In particular, \textit{the design of optimal periodic actuation inputs for soft robotic crawlers}, which similarly to inchworms, move by a combination of  peristaltic waves propagating along their elastic bodies and anisotropic friction with the substrate. As historically most of the literature has focused on rigid robotics \cite{DellaSantina2023},  
a principled framework for gait design in the emerging field of soft robotics is lagging behind: open loop controllers for gait sequences are often hand-crafted, which is non-trivial and time-consuming \cite{Ketchum2023}. To bridge this gap,
we propose two methods to design optimal periodic actuation inputs that generate crawling gaits for a two-segmented soft robotic crawler \cite{Arbelaiz_2024}. This crawler, as a modular building block, enables constructing complex multi-segmented crawlers \cite{Paoletti2014,Angatkina2023} and, more broadly, soft intelligent materials.

The contributions of this letter are twofold. We start by considering sinusoidal forcing profiles. Using describing function analysis \cite[Ch. 5]{Gelb1968,Slotine1991}, we prove that a \textit{matching condition is optimal:} 
the average center of mass speed is maximized when the forcing frequency matches the natural frequency of the crawler's body,  highlighting  \textit{the advantage of crawling at mechanical resonance}.  Building upon this result and extending  the formulation in \cite{Craun2015}, 
we propose a framework for \textit{gait optimization} based on Optimal Periodic Control (OPC) \cite{Colonious1976}. 
The cost objective in our proposed OPC problem trades off the speed of the crawler's center of mass with control effort and strain levels in the robot, important for the safe operation of the soft crawler.  We derive optimality conditions for the OPC problem using the calculus of variations and design a hill-climbing algorithm to solve it numerically. We illustrate the effectiveness of such a framework through a case study, where numerical results reveal a preference for crawling at an \textit{integer multiple of the natural frequency} when high control efforts and strains are penalized in the cost objective.

The letter is organized as follows. \S \ref{sec: model dynamics} introduces 
the model of the dynamics of the soft two-segmented crawler.
The problem of characterizing the optimal forcing frequency for average speed maximization subject to sinusoidal actuation is analyzed in \S \ref{sec:describing_function_analysis}. Some technical details are deferred to the \hyperlink{sec:Appendix}{Appendix} to streamline the presentation. \S \ref{sec: OPC} introduces our proposed OPC problem, the hill-climbing optimization algorithm to numerically solve it, and an illustrative case study. Conclusions are drawn in \S \ref{sec:conclusions}.

\section{Model of the crawler dynamics}
\label{sec: model dynamics}


\subsection{Crawler dynamics}
We utilize a minimal model of the dynamics of a soft two-segmented robotic crawler, resembling those presented in e.g., \cite{Arbelaiz_2024, Paoletti2014}. A schematic of the crawler is provided in Fig. \hyperref[Fig:friction_effect]{1\textbf{E}}. The crawler's dynamics are 
\begin{subequations}
    \begin{align}
        & m\Ddot{x}_1 = k(x_2-x_1)+b(\dot{x}_2-\dot{x}_1) + A_{\sigma} \sigma(\dot{x}_1) +A_f f,   
        \label{eq: dimensional dynamics 1}\\
        & m\ddot{x}_2 = k(x_1-x_2)+b(\dot{x}_1-\dot{x}_2)+A_{\sigma} \sigma(\dot{x}_2)-A_f f,
        \label{eq: dynamics 2}
    \end{align}
    \label{eq: dimensional dynamical equations}
\end{subequations}
\par\noindent where $x_i$ denotes the displacement of mass $i$ ($i = \{1, 2\}$) and the time dependence has been omitted for notational compactness. $\dot{x}_i$ and $\ddot{x}_i$ denote the velocity and acceleration of mass $i$, respectively. $k$ is the spring stiffness and $b$ is the damping constant. The anisotropic frictional force $\sigma(\cdot)$ enables net displacement of the crawler's center of mass (CoM). The unit amplitude actuation force $f$ (resp., frictional force $\sigma(\cdot)$) is scaled by the constant $A_f$ (resp. $A_\sigma$). 

\subsubsection{Anisotropic Friction Model}   

Soft-bodied crawling animals and robots leverage friction anisotropy to efficiently traverse various substrates.
We model the frictional force as a nonlinear anisotropic function of the crawler's local speed: the frictional force to move forward is small, as compared to that corresponding to backward motion. This defines the forward direction as the preferential direction of motion for the crawler. The frictional force $\sigma(\cdot)$ is modeled as an asymmetric sigmoid function of the local speed $\dot{x}_i$:
\begin{equation}
    \sigma(\dot{x_i}) =  \frac{1}{2}(\frac{1+n_f}{1+e^{-(-\dot{x}_i - \dot{x}_{\text{offset}})/\varepsilon_f}}+1-n_f),
\label{eq: friction}
\end{equation}
where the parameter 
$n_f$ tunes the friction anisotropy, $\varepsilon_f$ defines the slope of the nonlinearity, and $\dot{x}_{\text{offset}}$ is chosen such that $\sigma(0) = 0$.

 




\subsection{Non-dimensional dynamics}


We non-dimensionalize the crawler dynamics  \eqref{eq: dimensional dynamical equations}. 
We set the characteristic temporal ($t_*$) and spatial ($\ell_*$) scales of the dynamics using the undamped natural frequency of a linearized crawler ($\omega_n := \sqrt{2k/m}$) and its natural length (denoted by $\ell$), respectively:
\begin{equation}
    t_* := \frac{1}{\omega_n} \; \text{ and } \; \ell_* := \ell.
    \label{eq:characteristic_scales}
\end{equation}
The respective dimensionless\footnote{We use $\mathsf{sans \; serif}$ font to represent dimensionless variables.} temporal coordinate and displacement are defined as 
$
    \mathsf{t} := \omega_n t \text{ and }  \mathsf{x} := x/ \ell.
    \label{eq:dimensionless_coordinates}
$
Accordingly, the dimensionless counterpart of the crawler's dynamics \eqref{eq: dimensional dynamical equations} are 
\begin{small}
\begin{subequations}
    \begin{align}
        & \mathsf{x}_1''  = \pi_\sigma \sigma(\mathsf{x}_1') + \frac{1}{2}(\mathsf{x}_2-\mathsf{x}_1) + \zeta (\mathsf{x}_2'-\mathsf{x}_1') + \pi_\mathsf{f} \, \mathsf{f},   
        \label{eq: dynamics 1}\\
        & \mathsf{x}_2'' =   \pi_\sigma \sigma(\mathsf{x}_2') + \frac{1}{2}(\mathsf{x}_1-\mathsf{x}_2) + \zeta (\mathsf{x}_1'-\mathsf{x}_2')-\pi_\mathsf{f} \, \mathsf{f},
        \label{eq: dynamics 2}
    \end{align}
    \label{eq:ND_dynamical equations}
\end{subequations}
\end{small}
\par\noindent where $(\cdot)' := d(\cdot)/d \mathsf{t}$ and the \textit{dimensionless groups} $\pi_\mathsf{f}, \pi_\sigma$ and $\zeta$ are defined as 
\begin{equation}
    \pi_\mathsf{f}  :=  \frac{A_f}{2k \ell}, \; \pi_\sigma  := \frac{A_\sigma}{2k\ell}, \text{ and } \zeta  := \frac{b}{\sqrt{2km}}. 
    \label{eq:dimensionless_groups}
\end{equation}
$\zeta$ is commonly referred to as the \textit{damping ratio}.\\

The objective for the remainder of this paper is to determine actuation force inputs generating crawling gaits according to different optimality criteria. We start by analyzing sinusoidal forcings and show that when the forcing frequency matches the undamped natural frequency $\omega_n$ of the crawler's body, the average speed of its CoM is maximized.

\section{Crawling at Resonance is Optimal}
\label{sec:describing_function_analysis}


Under the assumption that the actuator force is sinusoidal, we aim to find the value of the forcing frequency that yields the maximum average speed in the crawler's CoM. We denote such an optimal forcing frequency by $\omega_*$.  For this analysis, in what follows, we approximate the response of the nonlinear crawler dynamics to the sinusoidal forcing through its \textit{describing function}. Describing function analysis, an extended version of the frequency response technique for linear systems, is a quasi-linearization method useful for predicting the response of a nonlinear system to a sinusoidal input -- see e.g., \cite[Ch. 5]{Gelb1968,Slotine1991}.
The core idea underlying the method is the approximation of the output of the nonlinear terms in the dynamics by an expansion in the \textit{fundamental harmonic},
ignoring higher-order harmonics as they are attenuated by the low-pass filtering structure of the plant, see Fig. 
\hyperref[Fig:friction_effect]{1\textbf{D}}. 

\subsection{Piecewise constant friction model}

\begin{figure}
\includegraphics[width=0.48\textwidth]{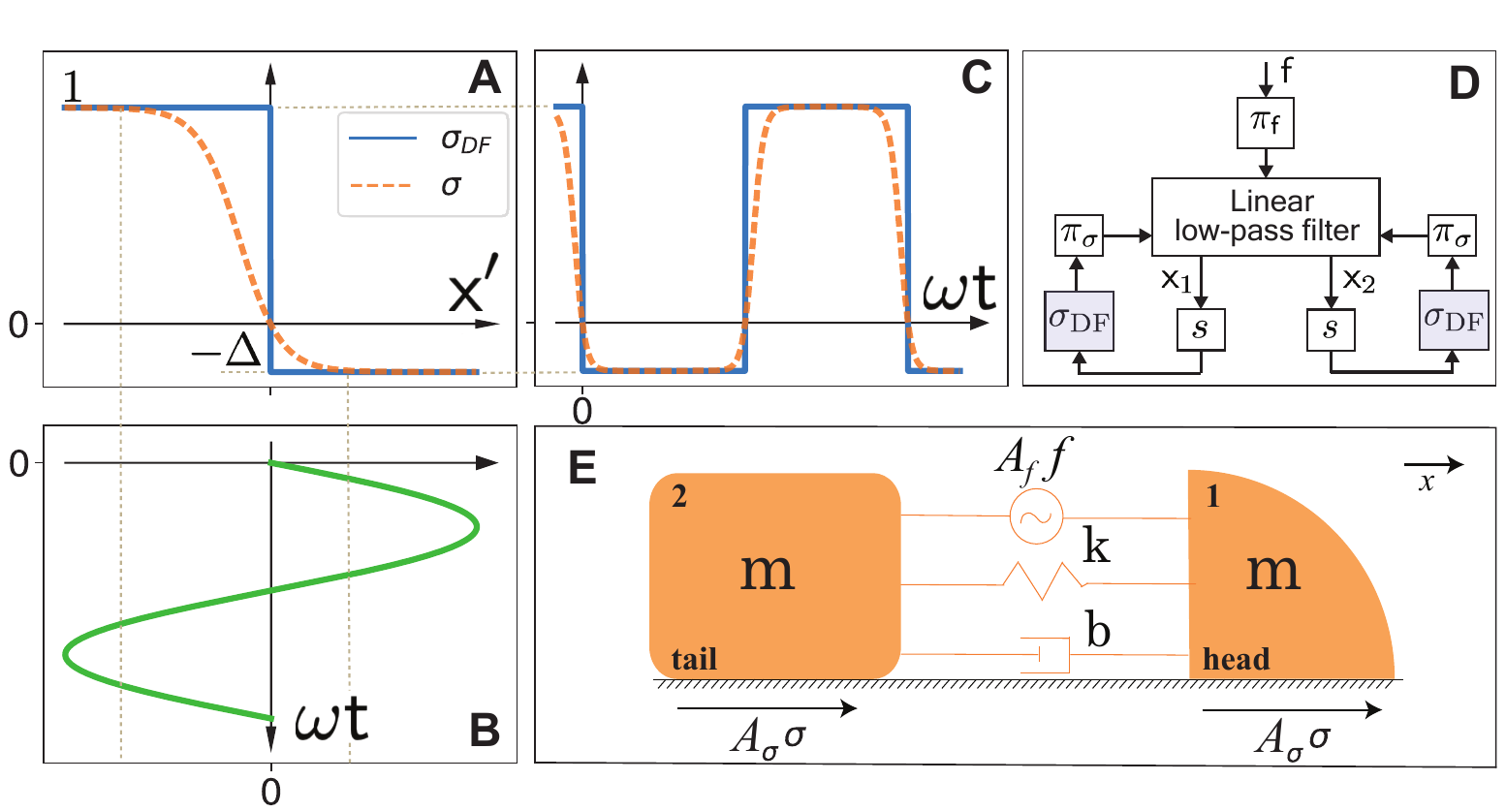}
\centering
\label{Fig:friction_effect}
\vspace{-3mm}
\caption{Nonlinear anisotropic friction models in the crawler dynamics. (\textbf{A}) Smooth ($\sigma$) and piece-wise constant ($\sigma_{\text{DF}}$) anisotropic friction input-output maps, as defined in \eqref{eq: friction} and \eqref{eq: piecewise constant friction (negative)}, respectively. The piece-wise constant model $\sigma_{\text{DF}}$ is used in the describing function analysis of Section~\S\ref{sec:describing_function_analysis} for the sake of analytical tractability. (\textbf{B}) Sinusoidal speed input. Horizontal axis in panels \textbf{A} and \textbf{B} are the same.
(\textbf{C}) Respective outputs of the friction models in panel \textbf{A} when the input is a sinusoidal local speed --- as depicted in \textbf{B}. Input and output are in phase. Color-code is the same in panels \textbf{A} and \textbf{C}.
(\textbf{D}) Block diagram of the system \eqref{eq:ND_dynamical equations}. Blocks shaded in blue capture the nonlinearity in the dynamics of the soft crawler. The low-pass filtering structure of the linear component of the crawler dynamics makes describing function analysis a suitable technique.   $s$ denotes the Laplace transform variable and the $s$-block corresponds to temporal differentiation. 
(\textbf{E}) Schematic of the soft crawler and its different components.
}

\end{figure}

In our describing function analysis and for the sake of analytical tractability, we approximate the sigmoidal friction model \eqref{eq: friction} by the following piecewise-constant model:
 \begin{small}
\begin{equation}
\label{eq: piecewise constant friction (negative)}
\sigma_{\text{DF}}(\dot{\mathsf{x}}_i) =
    \begin{cases}
      -\Delta, & \dot{\mathsf{x}}_i\geq 0,\\
      1, & \dot{\mathsf{x}}_i < 0,
    \end{cases}  
\end{equation}
\end{small}


where the parameter $0 < \Delta \ll 1$ defines the friction anisotropy and relates to the continuous friction as $\Delta = \frac{1}{2}(n_f - 1)$.
The approximation to the sigmoidal friction model and its output for a sinusoidal input are depicted in Fig. \hyperref[Fig:friction_effect]{1\textbf{A}} and Fig. \hyperref[Fig:friction_effect]{1\textbf{C}}, respectively. 

\subsection{The matching condition: crawling at resonance}

The following proposition states the main result of this section: when forced by a sinusoid, forcing at the undamped natural frequency of the crawler produces maximal average CoM speed.

\begin{prop}[Optimal frequency for sinusoidal forcing]
\label{prop:forcing_frequency}
Consider the non-dimensional crawler dynamics \eqref{eq:ND_dynamical equations} with the piecewise constant friction model \eqref{eq: piecewise constant friction (negative)} forced by a sinusoidal input $\mathsf{f}$.
Then, the describing function approximation establishes that the maximal average center of mass speed, $\bar{\nd{v}}_{\nd{com}}^*$, is achieved at the forcing frequency $\omega_* = 1$ and takes the value
\begin{equation}
    \bar{\nd{v}}_{\nd{com}}^* = \cos\bigg( \frac{\Delta \pi}{1+\Delta}\bigg)\frac{\pi_\nd{f} - 2\pi_\sigma \frac{1+\Delta}{\pi} \sin\big(\frac{\Delta \pi}{1 + \Delta}\big)}{2 \zeta}.
    \label{eq:max_average_vcom}
\end{equation}
\end{prop}
\begin{proof} It follows using describing function analysis and harmonic balance.  The main steps and intermediate results are provided here. Computation details are deferred to the \hyperlink{sec:appendix}{Appendix}. 

\textit{Harmonic approximations.} The actuation input in \eqref{eq:ND_dynamical equations} is set to
$
    \mathsf{f}(\mathsf{t}) = \sin(\omega \mathsf{t}+ \phi),
    \label{eq:sin_forcing}
$
where $\phi$ denotes the phase with respect to crawler's strain $\mathsf{s}$: 
\begin{equation}
    \mathsf{s}(\mathsf{t}) := \mathsf{x}_1(\mathsf{t}) - \mathsf{x}_2(\mathsf{t}) \approx \mathsf{A}\sin(\omega \mathsf{t}).
    \label{eq: strain equation}
\end{equation}
Without loss of generality, we take $\nd{A}>0$. The speeds $\mathsf{x}_i'$ 
are assumed shifted sinusoidals 
\begin{subequations}
    \begin{align}
        & \mathsf{x}_1'(\mathsf{t})  := \Bar{\mathsf{v}}_1 + \Tilde{\mathsf{v}}_1\cos(\omega \mathsf{t} +\theta_1),  \\
        & \mathsf{x}_2'(\mathsf{t})  := \Bar{\mathsf{v}}_2 +\Tilde{\mathsf{v}}_2\cos(\omega \mathsf{t} + \theta_2),
    \end{align}
    \label{eq: velocity form}
\end{subequations}
\vspace{-14pt}
\par\noindent where 
$\theta_i$ is the phase of the respective nodal speed with respect to the strain, $\bar{\mathsf{v}}_i>0$ is the average nodal speed, and $\tilde{\mathsf{v}}_i>0$ are coefficients. 

\textit{Fundamental harmonic of the nonlinear friction.}
For speeds of the form \eqref{eq: velocity form}, following \cite[Ch. 5]{ Gelb1968,Slotine1991} the frictional force \eqref{eq: piecewise constant friction (negative)} is approximated by  
\begin{equation}
    \sigma_{\text{DF}}(\mathsf{x}_i') \approx  \sigma_{\text{DF}}^{(0)}(\mathsf{x}_i') + \sigma_{\text{DF}}^{(1)}(\mathsf{x}_i')\cos(\omega \mathsf{t}) +\sigma_{\text{DF}}^{(2)}(\mathsf{x}_i') \sin(\omega \mathsf{t}),
    \label{eq:describing_fcn_friction}
\end{equation}
where $\sigma_{\text{DF}}^{(0)}$, $\sigma_{\text{DF}}^{(1)}$, and $\sigma_{\text{DF}}^{(2)}$ are the Fourier series coefficients for the DC term and the fundamental harmonic, respectively, given by\footnote{
The analytical computation of the coefficients in \eqref{eq:friction_coeffs} is in Appendix \ref{subsec:friction_coeffs_harmonic_balance}.} 
\begin{small}
\begin{subequations}
    \begin{align}
        \sigma_{\text{DF}}^{(0)}(\mathsf{x}_i') 
        &  = \frac{1}{\pi}(-\Delta\pi+(1+\Delta)\cos^{-1}(\mathsf{a}_i)), \label{eq: friction eq 1}\\
        \sigma_{\text{DF}}^{(1)}(\mathsf{x}_i') 
        & = -\frac{2(1+\Delta)}{\pi }\cos(\theta_i)\sqrt{1-\mathsf{a}_i^2}, \label{eq: friction eq 2}\\
        \sigma_{\text{DF}}^{(2)}(\mathsf{x}_i') 
        & = \frac{2(1+\Delta)}{\pi }\sin(\theta_i)\sqrt{1-\mathsf{a}_i^2}, \label{eq: friction eq 3}
    \end{align}
    \label{eq:friction_coeffs}
\end{subequations}
\end{small}
\par\noindent where $ 0< a_i := \bar{\mathsf{v}}_i/\tilde{\mathsf{v}}_i <1, \text{ with } i = 1,2$. 






\textit{Harmonic balance.} Harmonic balance\footnote{The detailed harmonic balance derivation is in Appendix \hyperref[subsec:harmonic_balance_appendix]{B}.} in \eqref{eq: strain equation}-\eqref{eq: velocity form} yields $\theta_1 = 0$ and $\theta_2 = \pi$. The variables $\phi$ and $\nd{A}$ must jointly satisfy
\begin{small}
\begin{subequations}
    \begin{align}
        \sin \phi & = \frac{1}{\pi_{\nd{f}}}\Big(\zeta \mathsf{A} \omega + 2 \pi_{\sigma}\frac{1+\Delta} {\pi}\sqrt{1-\mathsf{a}^2}\Big), \label{eq: phi A relation 1}\\
        \cos \phi & = \frac{\mathsf{A}}{2\pi_{\nd{f}}}  (1-\omega^2). \label{eq: phi A relation 2}
    \end{align}
        \label{eq:phi_and_A}
\end{subequations}
\end{small}
\par\noindent Furthermore, $\bar{\nd{v}}_1 = \bar{\nd{v}}_2 =:\bar{\nd{v}}$, given by
\begin{small}
\begin{equation}
    \Bar{\mathsf{v}} = \frac{\omega \mathsf{A} \nd{a}}{2}. \label{eq: V_bar A relation main}
\end{equation}
\end{small}


\textit{Optimal forcing frequency.}
\label{subsec: optimal omega}
We denote the average speed of the crawler's CoM by $\bar{\mathsf{v}}_{\text{com}}$.
We aim to find the forcing frequency $\omega_*$ maximizing $\bar{\mathsf{v}}_{\text{com}}$.
From the previous analysis, 
\begin{equation}
    \bar{\mathsf{v}}_{\text{com}} = \bar{\mathsf{v}} = \frac{\omega \mathsf{A} \nd{a}}{2} = \frac{\nd{a}}{2 \, \zeta} \big(\pi_{\nd{f}} \sin\phi-2\pi_{\sigma}\frac{1+\Delta}{\pi}\sqrt{1-\mathsf{a}^2}\,\big),
    \label{eq:v_com_main}
\end{equation}
where the second equality follows from \eqref{eq: V_bar A relation} and the third equality from rewriting \eqref{eq: phi A relation 1}. Stationary points $\bar{\omega}$ satisfy
\begin{equation*}
    \frac{d \, \bar{\mathsf{v}}_{\text{com}}}{d \, \omega} \bigg \vert_{\bar{\omega}} = 0 \; \Rightarrow \; \frac{\nd{a}\,  \pi_{\nd{f}}}{2 \zeta} \cos(\phi) \frac{d \, \phi}{d \, \omega}\bigg\vert_{\bar{\omega}} = \frac{\nd{a}\, \mathsf{A}}{4 \zeta }(1- \bar{\omega}^2) \frac{d \, \phi}{d \, \omega}\bigg\vert_{\bar{\omega}}  = 0,
\end{equation*}
where we used \eqref{eq: phi A relation 2}, yielding $\bar{\omega} = 1$. 
Furthermore,
\begin{equation*}
    \frac{d^2 \, \bar{\mathsf{v}}_{\text{com}}}{d \, \omega^2}   = \frac{\nd{a}\,\pi_{\nd{f}}}{2\zeta}\Big(-\sin(\phi) \Big(\frac{d \, \phi}{d \, \omega}\Big) ^{2}+\cos{\phi}\frac{d^2 \, \phi}{d \, \omega^2}\Big).
\end{equation*}
From \eqref{eq: phi A relation 2}  at $\bar{\omega} = 1, \; \cos{\phi} = 0$ 
and therefore, $\sin{\phi} = \pm 1$. Under the assumption $\mathsf{A}>0$, then from \eqref{eq: phi A relation 1}  $\sin{\phi}>0$. Thus, $\frac{d^2 \, \bar{\mathsf{v}}_{\text{com}}}{d \, \omega^2} \big \vert_{\bar{\omega}} < 0 \;\; \Rightarrow \;\; \omega_* = \bar{\omega} =  1$ is the forcing frequency producing \textit{maximal} $\bar{\mathsf{v}}_{\text{com}}$, denoted by $\bar{\mathsf{v}}_{\text{com}}^*$. Substitution of these values into \eqref{eq:v_com_main} yields $\bar{\mathsf{v}}_{\text{com}}^*$ in \eqref{eq:max_average_vcom}. 
\end{proof}
We highlight two interesting observations related to the result in Proposition \ref{prop:forcing_frequency}:
\begin{itemize}
    \item $\omega_*= 1$ is the optimal dimensionless forcing frequency. In dimensional frequency, it corresponds to a forcing frequency of $\omega_n$, the \textit{undamped natural frequency} of the crawler. This defines a \textit{matching condition} \cite{Simoni2007,Arbelaiz2023, Arbelaiz2020, Arbelaiz2022}: for the crawler to maximize its average CoM speed when subject to a sinusoidal actuation force, the forcing frequency must \textit{match} the natural frequency of the crawler's body.
    Interestingly, further advantages of moving at mechanical resonance have been reported in the literature, such as minimum metabolic cost, reduction of jitter during movements, and improvement of cycle-to-cycle reproducibility \cite{Simoni2007,Goodman2000}.
    \item The \textit{instantaneous actuation power}  is defined as
    $
        \mathcal{P}_{\omega}(\nd{t}):= \nd{f}(\nd{t}) \, \nd{s}'(\nd{t}), \; \nd{t}\geq 0,
        \label{eq:instantaneous_power_def}
    $
    where the subindex $\omega$ indicates its dependency on the forcing frequency. $\mathcal{P}_\omega$ is the power that the actuator provides to the two crawling segments. By \eqref{eq:phi_and_A}, at $\omega_* = 1$, $\phi = \frac{\pi}{2} \; \Rightarrow
    \; \nd{f}(\nd{t}) = \nd{s}'(\nd{t}) = \nd{A} \cos{(\nd{t})}$. Thus, the optimal actuation input and the corresponding strain rate are \textit{in phase}. This implies that $\mathcal{P}_{\omega_*} = \nd{A}^2 \cos^2(\nd{t})$ is a positive function, which generally does not hold for an arbitrary forcing frequency $\omega$ due to a phase difference between forcing and strain rate. The physical interpretation of such positivity is that at $\omega_*=1$  the actuator does not extract power from the 
 two segments of the crawler.    
\end{itemize}

Through describing function analysis, in this section we have established that forcing the crawler at mechanical resonance is optimal to maximize the average speed of its center of mass. However, there might be other performance metrics of interest for the successful operation of the soft robot:  it might be desirable to penalize high strain amplitudes to preserve the integrity of the robotic platform and to avoid high control efforts. We address the design of optimal crawling gaits accounting for a trade-off of these additional performance criteria in \S\ref{sec: OPC}.

\section{Optimal Periodic Control: 
Gait Optimization}
\label{sec: OPC}
We propose a methodology based on \textit{Optimal Periodic Control (OPC)} \cite{Colonious1976} to find \textit{periodic} exogenous input force profiles leading to optimal periodic crawling gaits. The performance criterion of the OPC problem trades off the maximization of the crawler's displacement with the minimization of strain amplitude and actuation effort. The force profile is enforced to be periodic, but not necessarily harmonic, generalizing the framework of \S \ref{sec:describing_function_analysis}. The optimal harmonic solution found in \S \ref{subsec: optimal omega} is utilized to warm start our proposed optimization algorithm, which numerically solves the OPC. Our problem formulation builds upon that introduced in \cite{Craun2015}, extending it to account for mixed boundary conditions. 

\subsection{Problem Formulation}
\label{subsec:OPC_formulation}

We define the state $\boldsymbol{\nd{z}}(\nd{t})$ of the crawler dynamics \eqref{eq:ND_dynamical equations} as $\boldsymbol{\nd{z}} := [ (\mathsf{x}_1 - \mathsf{x}_2) \;\; (\mathsf{x}_1+\mathsf{x}_2) \;\; \mathsf{x}_1' \;\;  \mathsf{x}_2']^{\top}$, which evolves according to
\begin{small}
\begin{equation}
\label{eq: state dynamics}
    \boldsymbol{\mathsf{z}'} = 
    \begin{bmatrix}
       \mathsf{z}_3-\mathsf{z}_4\\
       \mathsf{z}_3+\mathsf{z}_4\\
       \;\; \, \pi_\sigma \sigma(\mathsf{z}_3)- 0.5 \, \mathsf{z}_1 - \zeta(\mathsf{z}_3-\mathsf{z}_4) + \pi_\nd{f}\mathsf{f}\\  
       \;\;\, \pi_\sigma \sigma(\mathsf{z}_4)+ 0.5 \, \mathsf{z}_1 + \zeta(\mathsf{z}_3-\mathsf{z}_4) -\pi_\nd{f} \mathsf{f}
    \end{bmatrix} =: \bold{g}(\boldsymbol{\nd{z}}, \nd{f}).
\end{equation} 
\end{small}

The entries of $\boldsymbol{\nd{z}}$ correspond to the crawler's strain, twice the total displacement of its center of mass, and the local nodal speeds, respectively. For a time-periodic input $\nd{f}$, the state entries ($\nd{z}_1, \nd{z}_3, \nd{z}_4$) are expected to be periodic.
The friction model $\sigma(\cdot)$ is as given in \eqref{eq: friction}.

The optimization criterion\footnote{The time-horizon $\nd{T}$ in the OPC problem is set by the designer. We use the notation  $\mathcal{J}_\nd{T}$ to highlight the parameterization of the cost functional by $\nd{T}$. Due to the periodic boundary conditions in the OPC problem, $\nd{T}$ is either the crawling period or an integer multiple of it.} $\mathcal{J}_{\nd{T}}$ trades off speed maximization with the minimization of actuator effort and strain in the crawler over a fixed time-horizon $\nd{T}$. The \textit{OPC problem} is 
\begin{small}
\begin{maxi}|s|[0]                     
{\nd{f}(\cdot),  \boldsymbol{\nd{z}}(\cdot)}
{
\begin{aligned}
     \overbrace{\frac{1}{\nd{T}} \Big( \mathsf{z}_2(\nd{T}) 
    - \alpha \int_0^{\nd{T}} \mathsf{f}(\nd{t})^2 d\mathsf{t} 
    - \beta \int_0^\nd{T} \mathsf{z}_1(\nd{t})^2 d\mathsf{t} \Big)}^{\mathcal{J}_{\nd{T}}(\nd{f}, \boldsymbol{\nd{z}})}
\end{aligned}
} 
{\label{optimizationProblem}} 
{} 
\addConstraint{\boldsymbol{\nd{z}}'=}{\; \mathbf{g}(\boldsymbol{\nd{z}}, \nd{f})}{} 
\addConstraint{\mathsf{z}_i(\nd{T}) = }{\; \mathsf{z}_i(0)}{}, \; i = 1,3,4
\addConstraint{\mathsf{z}_2(0) = }{\; 0}{} 
\addConstraint{\mathsf{f}(\nd{T}) = }{\; \mathsf{f}(0)}{}, 
\end{maxi}
\end{small}
\par \noindent where\footnote{We define the function space $\mathcal{C}^1([a,b], \mathbb{R}^n):= \{ \boldsymbol{f}:[a,b] \to \mathbb{R}^n: \boldsymbol{f} \text{ is continuously differentiable}\}$, equipped with the norm $\| \boldsymbol{f}\|_1 := \max_{t \in [a,b]} \| \boldsymbol{f}(t)\| + \max_{t \in [a,b]} \| \boldsymbol{f}'(t)\|\}$, where $\| \cdot \|$ denotes the standard Euclidean norm in $\mathbb{R}^n$. 
} \big($\nd{f}(\cdot),  \boldsymbol{\nd{z}}(\cdot)\big) \in \mathcal{C}^1([0,\nd{T}], \mathbb{R}^5)$; $\alpha, \beta \in \mathbb{R}_+$ are relative penalization parameters among the different terms in the cost functional, their value being set by the designer;
   the dynamics of the state $\boldsymbol{\nd{z}}(\nd{t})$ are as defined in \eqref{eq: state dynamics}; and
    $\mathsf{f}, \, \mathsf{z}_1$, $\mathsf{z}_3$, and $\mathsf{z}_4$ are subject to periodic boundary conditions.
    Without loss of generality, the initial condition on the CoM displacement is set to $\mathsf{z}_2(0) = 0$. 


\subsection{Optimality conditions: first variation \& co-state dynamics}

Calculating the first variation of the Lagrangian is essential for deriving first-order optimality conditions and numerical algorithms. We define the Lagrangian functional $\mathcal{L}_\nd{T}$
associated with the constrained functional optimization \eqref{optimizationProblem} by introducing the Lagrange multiplier function $\boldsymbol{\lambda}: [0, \nd{T}] \to \mathbb{R}^4$, 
 usually referred to as the \textit{adjoint vector} or \textit{co-state}  \cite[\S 2.5.2]{Craun2015, liberzon2011calculus}:  
\begin{small}
\begin{align}
\mathcal{L}_\nd{T}(\boldsymbol{\nd{z}}, \nd{f}, \boldsymbol{\lambda}) & := \frac{1}{\nd{T}} \bigg( \mathsf{z}_2(\nd{T}) - \alpha\int_0^\nd{T} \mathsf{f}(\nd{t})^2 d\mathsf{t} -\beta\int_0^\nd{T} \mathsf{z}_1(\nd{t})^2 d\mathsf{t}  \nonumber \\ 
& \hspace{1.75cm} - \int_0^\nd{T} \boldsymbol{\lambda}(\nd{t})^\top \big(\boldsymbol{\mathsf{z}}'(\nd{t})-\bold{g}(\boldsymbol{\nd{z}}, \nd{f})\big)d \mathsf{t}\bigg).
    \label{eq:Lagrangian}
\end{align}
\end{small}
After integration by parts, the first variation $ \delta \mathcal{L}_\nd{T}$ of the Lagrangian $\mathcal{L}_\nd{T}$ defined in \eqref{eq:Lagrangian} is
\begin{small}
\begin{align}
    \delta \mathcal{L}_\nd{T} & = \frac{1}{\nd{T}} \bigg( \delta \mathsf{z}_2(\nd{T}) 
    - 2\alpha \int_0^\nd{T} \mathsf{f}(\nd{t}) \, \delta \mathsf{f}(\nd{t}) d\mathsf{t}
     -2\beta \int_0^\nd{T} \mathsf{z}_1(\nd{t}) \delta \mathsf{z}_1(\nd{t}) d\mathsf{t} \nonumber\\
     & - \big[\boldsymbol{\lambda}(\nd{t})^\top \delta \boldsymbol{\mathsf{z}}(\nd{t})\big]_0^\nd{T} +\int_0^\nd{T}\boldsymbol{\lambda}'(\nd{t})^\top \delta \boldsymbol{\mathsf{z}}(\nd{t}) d \mathsf{t}+\int_0^\nd{T} \boldsymbol{\lambda}(\nd{t})^\top \frac{\partial \bold{g}}{\partial \boldsymbol{\mathsf{z}}} \, \delta \boldsymbol{\mathsf{z}}(\nd{t})    d\mathsf{t} \nonumber \\
    & + \int_0^\nd{T} \boldsymbol{\lambda}(\nd{t})^\top \frac{\partial \bold{g}}{\partial \mathsf{f}} \, \delta \mathsf{f}(\nd{t}) d\mathsf{t}\bigg),
    \label{eq:lagrangian_first_variation}
\end{align}
\end{small}
\par \noindent where $\delta \nd{f}(\cdot)\in \mathcal{C}^1([0,\nd{T}], \mathbb{R})$ 
and $\delta \boldsymbol{\nd{z}}(\cdot) \in \mathcal{C}^1([0,\nd{T}], \mathbb{R}^4)$ 
denote perturbations to $\nd{f}(\cdot)$ and $\boldsymbol{\nd{z}}(\cdot)$, respectively. In order for these perturbations to be \textit{admissible}, they must satisfy the following \textit{periodic boundary conditions} 
        $
        \delta \mathsf{f}(\nd{T}) = \delta \mathsf{f}(0) \text{ and }
        \delta \mathsf{z}_{1, 3, 4}(\nd{T}) = \delta \mathsf{z}_{1, 3, 4}(0)
        $
-- making the perturbations $\delta \nd{f}, \delta \nd{z}_1, \delta \nd{z}_3 \text{ and } \delta \nd{z}_4$ periodic -- 
and the initial condition $\delta \mathsf{z}_2(0) = 0$.

At an extremal, the first variation \eqref{eq:lagrangian_first_variation} must vanish. Using the Fundamental Lemma of the Calculus of Variations \cite[a mild modification of Lemma 2.1] 
{liberzon2011calculus} 
yields the \textit{co-state dynamics}
\begin{subequations}
\begin{small}
\begin{equation}
\label{eq: costate dynamics}
    \boldsymbol{\lambda}' = 
    \begin{bmatrix}
        0& 0& \frac{1}{2} & -\frac{1}{2}\\
        0& 0&  0&0\\
        -1&  -1& \zeta -\pi_\sigma \sigma'(\nd{z}_3) &- \zeta \\
        1 &-1& -\zeta& \zeta-\pi_\sigma \sigma'(\nd{z}_4)
    \end{bmatrix} \boldsymbol{\lambda}+ \begin{bmatrix}
        2 \beta\mathsf{z}_1\\
        0\\
        0 \\
        0 
    \end{bmatrix},
\end{equation}
\end{small}
\par \noindent and the following identity among boundary terms 
$
    \big[\delta \nd{z}_2(\nd T) - \boldsymbol{\lambda}(\nd t)^\top \delta \boldsymbol{\mathsf{z}}(\nd t)\big]^\nd{T}_0 = 0,
    \label{eq:identity at boundary}
$
which in turn provides the \textit{boundary conditions} for the co-state dynamics
\begin{equation}
\label{eq: costate dynamics bc}
    \begin{aligned}
        \lambda_i(\nd{T}) = \lambda_i(0) \text{ with } i = 1,3,4 \text{ and }
        \lambda_2(\nd{T}) = 1.
    \end{aligned}
\end{equation}
\label{eq:costate_dynamics_all}
\end{subequations}
\par\noindent \eqref{eq: costate dynamics} and \eqref{eq: costate dynamics bc} together  $\Rightarrow \, \lambda_2(\nd{t}) = 1, \,  \nd{t} \in [0,\nd{T}].$ The dynamics \eqref{eq: state dynamics} and \eqref{eq: costate dynamics} with their respective initial and boundary conditions provided in \eqref{optimizationProblem} and \eqref{eq: costate dynamics bc} form a set of coupled nonlinear ODEs with \textit{mixed boundary conditions} composed by six \textit{two-point boundary value problems} (TPBVPs) and one \textit{initial value problem} (IVP) -- as $\lambda_2$ has already been solved for. The boundary conditions in the TPBVPs are periodic, but their values are unknown and to be determined.
Using \eqref{eq: costate dynamics} the first  variation \eqref{eq:lagrangian_first_variation} simplifies to
\begin{small}
\begin{equation}
    \delta \mathcal{L}_{\nd{T}} = \frac{1}{\nd{T}} \int_0^\nd{T} \bigg[ \boldsymbol{\lambda}(\nd{t})^\top \frac{\partial \bold{g}}{\partial \nd{f}} (\boldsymbol{\nd{z}}, \nd{f}) - 2 \alpha \nd{f}(\nd{t}) \bigg] \delta \nd{f}(\nd{t}) d \nd{t}.
    \label{eq:first_variation_simplified}
\end{equation}
\end{small}
 \par \noindent The bracketed term in the integrand of \eqref{eq:first_variation_simplified} is set to zero to complete the first order necessary conditions for optimality, providing an expression for $\nd{f}(\cdot)$ in terms of the Lagrange multiplier $\boldsymbol{\lambda}(\cdot)$. Alternatively, we use \eqref{eq:first_variation_simplified} to propose an expression for the update $\delta \nd{f}(\cdot)$ of the control input to be utilized in a numerical optimization scheme, presented next.



\subsection{
Hill-climbing algorithm for the OPC problem}
\begin{algorithm}[b]
\caption{Hill-climbing for the OPC problem}\label{alg:hillClimbing_OPC}
\begin{algorithmic}
\Require Initialize $\mathsf{f}_0$ (according to Proposition \ref{prop:forcing_frequency})
\While{not converged}

    \State $\boldsymbol{\mathsf{z}}_n(\cdot) \gets$ Use direct collocation in \eqref{eq: state dynamics} and solve
    \begin{equation*}
        \setlength{\jot}{-3 pt} 
        \begin{aligned}
        \indent \min_{\boldsymbol{\mathsf{z}}_n(\cdot)} \quad & 1\\
        \textrm{s.t.} \quad & \eqref{eq: state dynamics} \text{ met on all collocation points,}\\
        & \nd{z}_{n,i}(0) = \nd{z}_{n,i}(\nd{T})\, (i = 1,3,4),\; \nd{z}_{n,2}(0) = 0.
        \end{aligned}
    \end{equation*}
    
    \State $\boldsymbol{\lambda}_n(\cdot) \gets$ Use direct collocation in  \eqref{eq:costate_dynamics_all} and solve
    \begin{equation*}
        \setlength{\jot}{-3 pt} 
        \begin{aligned}
        \indent \min_{\boldsymbol{\lambda}_n(\cdot)} \quad & 1\\
        \textrm{s.t.} \quad & \eqref{eq: costate dynamics} \text{ met on all collocation points,}\\
        & \nd{\lambda}_{n,i}(0) = \nd{\lambda}_{n,i}(\nd{T})\, (i = 1,3,4),\; \lambda_{n,2}(\nd{T}) = 1.
        \end{aligned}
    \end{equation*}
    
    \State $\mathsf{f}_{n+1} \gets$ $\mathsf{f}_{n}+\epsilon\, \delta \mathsf{f}_n$
\EndWhile

\Return $\mathsf{f}_{n+1}$, $\boldsymbol{\mathbf{\nd{z}}}_{n+1}, \boldsymbol{\lambda}_{n+1}$
\end{algorithmic}
\end{algorithm}
We utilize \eqref{eq:first_variation_simplified} to build an algorithm \footnote{Code at \href{https://github.com/Danielshen2000/OptimalCrawlerGait}{\texttt{github.com/Danielshen2000/OptimalCrawlerGait}}} to maximize the objective $\mathcal{J}_\nd{T}$ (and thus, the term ``hill-climbing''). We set
\begin{small}
\begin{equation}
    \delta \nd{f}(\nd{t})  = \boldsymbol{\lambda}(\nd{t})^\top \frac{\partial \bold{g}}{\partial \nd{f}} (\boldsymbol{\nd{z}}, \nd{f}) - 2 \alpha \nd{f}(\nd{t})    = \pi_\nd{f} \big(\lambda_3(\nd{t}) -\lambda_4(\nd{t})\big) - 2 \alpha \nd{f}(\nd{t}),
    \label{eq:delta_f}
\end{equation}
\end{small}
\par \indent such that $\delta \mathcal{L}_{\nd{T}} \geq 0$. Our hill climbing algorithm repeatedly updates the actuation input $\mathsf{f}(\cdot)$ using the $\delta \mathsf{f}(\cdot)$ given in \eqref{eq:delta_f}, as described next.  $(\boldsymbol{\nd{z}}_n, \nd{f}_n, \boldsymbol{\lambda}_n)$ denotes the value of the functions $(\boldsymbol{\nd{z}}, \nd{f}, \boldsymbol{\lambda})$ at the $n-$th iteration of the algorithm.
 The OPC problem \eqref{optimizationProblem} is recursively solved by iterating between two different steps until convergence 
 (see Algorithm \ref{alg:hillClimbing_OPC} for an overview of the numerical routine):
 \paragraph{Solving for periodic state and co-state trajectories:} 
 
 Given a actuation input $\nd{f}_n(\cdot)$, the state dynamics \eqref{eq: state dynamics} subject to the boundary conditions in \eqref{optimizationProblem} are solved for periodic trajectories as a joint TPVBPs and IVPs through a \textit{feasibility problem}. Once $\boldsymbol{\nd{z}}_n(\cdot)$ is obtained, the algorithm proceeds analogously with the co-state dynamics and boundary conditions in \eqref{eq:costate_dynamics_all}.
Direct collocation methods are utilized to approximate the ODEs with mixed boundary conditions by a set of algebraic equations. 
Specifically, the open-source Python package \texttt{Pyomo} \cite{hart2017pyomo} is used to discretize both the state and the co-state dynamics using the Lagrange-Radau transformation. The feasibility problem is numerically solved by interior point methods, using \texttt{IPOPT} \cite{wachter2006IPOPT}. 

\paragraph{Actuation update \& hill climbing:} The actuation input $\nd{f}(\cdot)$ is updated using \eqref{eq:delta_f}. The increment in the cost functional is guaranteed by choosing the stepsize $\epsilon$ (see Algorithm \ref{alg:hillClimbing_OPC}) sufficiently small \cite[\S 3.2]{Craun2015}.

\subsection{A case study}
\label{subsec:case_study}
We present a case study to illustrate the effectiveness of the OPC problem of \S\ref{subsec:OPC_formulation} and Algorithm \ref{alg:hillClimbing_OPC} in generating crawling gaits. 

\paragraph{System set-up:} We analyze a soft crawler whose characteristic scales \eqref{eq:characteristic_scales} are defined by a natural length of $\ell = 0.1\text{m}$ and an undamped natural frequency of $\omega_n = 0.4472$ rad/s. The dimensionless groups in the dynamics of the crawler \eqref{eq:ND_dynamical equations} take values $\pi_\sigma = \pi_\nd{f} = 1$ and $\zeta = 0.2236$, respectively. The relative penalization parameters of the cost functional $\mathcal{J}_\nd{T}$ in \eqref{optimizationProblem} are $\alpha = 3.3$ and $\beta = 0.05$. The time-horizon is $\nd{T} = 2\pi$.

\paragraph{Numerical optimization:}
Regarding the configuration of the numerical solver for the ODEs with mixed boundary conditions, the interval $[0,\nd{T}]$ is split into 30 sub-intervals for numerical evaluation, each sub-interval containing 10 collocation points where temporal derivatives are approximated. 
Algorithm \ref{alg:hillClimbing_OPC} is initialized at $\nd{f}_0(\nd{t}) = \sin{(\nd{t})}$ and the stepsize of the actuation update is set to $\epsilon = 0.01$.

\begin{figure}
\includegraphics[width=0.47\textwidth]{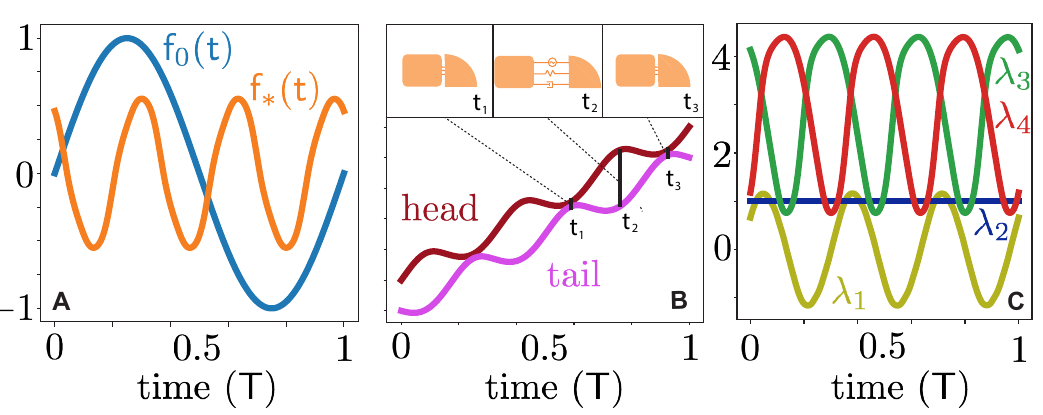}
\label{fig:optimization_result}
\centering
\caption{
Optimal solution to the OPC problem \eqref{optimizationProblem}, numerically obtained using Algorithm \ref{alg:hillClimbing_OPC}. Parameter values are as described in \S \ref{subsec:case_study}.
(\textbf{A}) Initial ($\nd{f}_0$) and optimized ($\nd{f}_*$) actuation force profiles in time. $\nd{f}_0$ is chosen according to Proposition \ref{prop:forcing_frequency}. 
Algorithm \ref{alg:hillClimbing_OPC} converges to a  periodic force profile, its period being an integer multiple of the crawler's natural frequency. 
(\textbf{B}) Positions of the soft crawler's head and tail in time, illustrating the optimal crawling gait. (\textbf{C}) Periodic trajectories of the co-states $\boldsymbol{\lambda}$, solution to \eqref{eq:costate_dynamics_all}. }
\end{figure}

\paragraph{Results:}
Fig. \hyperref[fig:optimization_result]{2} 
illustrates the efficacy of the OPC problem formulation and the corresponding Algorithm \ref{alg:hillClimbing_OPC} in generating crawling gaits. Algorithm \ref{alg:hillClimbing_OPC} increased the value of $\mathcal{J}_\nd{T}$ from $-17.08$ (corresponding to a  resonant sinusoidal initialization $\nd{f}_0$) to $2.43$ (converged solution). An interesting observation is that when the performance objective $\mathcal{J}_\nd{T}$ accounts for the strain and control effort, in addition to the average speed of the crawler's CoM, the numerical scheme converges to a periodic forcing with a frequency that is an integer multiple of the crawler's undamped natural frequency (that is, it resembles a \textit{higher order harmonic}) -- see Fig. \hyperref[fig:optimization_result]{2}\textbf{A}. 
This result indicates \textit{a preference for shorter crawling gaits when excessive force efforts and strain levels in crawler's body are penalized}.
Similar harmonic effects have been observed in the optimal control of quantum systems \cite{Grivopoulos2004}. The average frictional force over a cycle for the optimal periodic solution is zero, consistent with \eqref{eq: COM dynamics 1} in the harmonic balance analysis.

\section{Conclusion}
\label{sec:conclusions}
We analyzed the efficiency of different periodic actuation profiles in generating peristaltic waves in a two-segmented soft robotic crawler. Together with anisotropic friction, these are responsible for the crawler's locomotion. First, an harmonic forcing pattern was considered. Through describing function analysis, 
we proved that \textit{optimality is achieved at a matching condition}: when the frequency of the harmonic forcing matches the undamped natural frequency of the crawler's body, the average speed of its center of mass is maximized. This result is well-aligned with previous literature reporting various advantages of \textit{moving at mechanical resonance} \cite{Simoni2007,Goodman2000}. Second, building upon this result, we introduced a framework grounded on Optimal Periodic Control (OPC) to design optimal periodic forcing profiles which need \textit{not} be harmonic, i.e., the optimal forcing can adopt an arbitrary waveform. In addition to the average speed of the soft crawler's center of mass, the optimization objective in our OPC problem accounts for its strain levels and control effort, critical for the safe operation of the robot. Based on first order  optimality conditions derived from the calculus of variations, we proposed a hill-climbing algorithm to numerically solve the OPC problem and illustrated its effectiveness through a case study. In this framework, numerical results suggest a preference for periodic forcing profiles where the frequency is an integer multiple of the crawler's undamped natural frequency -- i.e., shorter crawling gaits.

On-going work includes analyzing the efficiency of different gaits in the soft robotic crawler  OSCAR \cite{Angatkina2023}. Extensions of the OPC problem include designing optimal peristaltic waves for multi-segmented soft crawlers and decentralized feedback controllers based on spatially invariant systems \cite{Bamieh,Epperlein2016,Arbelaiz2021}.




{\sc Acknowledgements}
The authors would like to thank Allison Earnhardt and Prof. Aimy Wissa (Princeton U.) for interesting discussions on the robotic platform OSCAR \cite{Angatkina2023}.

\appendix
\begingroup
\small
\setlength{\abovedisplayskip}{3pt} 
\setlength{\belowdisplayskip}{3pt} 
\label{sec:Appendix}
\numberwithin{equation}{subsection}
\subsection{Analytical computation of the fundamental harmonic of the frictional force }
\label{subsec:friction_coeffs_harmonic_balance}

We compute the Fourier coefficients \eqref{eq:friction_coeffs} in the approximation \eqref{eq:describing_fcn_friction} of the frictional model \eqref{eq: piecewise constant friction (negative)} when the speed is of the form \eqref{eq: velocity form}. Define $\mathsf{a}_i := \Bar{\mathsf{v}}_i/\Tilde{\mathsf{v}}_i >0$ by assumption and note that $
    \sigma_{\text{DF}}\big(\Bar{\mathsf{v}}_i+\Tilde{\mathsf{v}}_i \cos(\omega \mathsf{t} +\theta_i)\big) = \sigma_{\text{DF}}\big(\mathsf{a}_i+\cos(\omega \mathsf{t} +\theta_i)\big).$ 
 $\mathsf{a}_i > 1$ 
 does \textit{not} correspond to crawling, as the 
 frictional force
 should alternate sign during a crawling cycle. Thus, we restrict the harmonic balance analysis to the region of physical relevance $0 < \mathsf{a}_i \leq 1$, for which 
\begin{subequations}
\begin{align}
    \mathsf{a}_i+\cos(\omega \mathsf{t} +\theta_i) & < 0 \text{ for } \omega \mathsf{t} \in (\alpha_i, \beta_i), \\
    \mathsf{a}_i+\cos(\omega \mathsf{t} +\theta_i) & \geq 0 \text{ for } \omega \mathsf{t} \in [-\theta_i,\alpha_i] \cup [\beta_i, 2\pi-\theta_i],
\end{align}
\label{eq:useful_a_region}
\end{subequations}
\par\noindent with 
$
        \alpha_i := \pi - \cos^{-1}(\mathsf{a}_i) - \theta_i 
    \text{ and } \beta_i := \pi + \cos^{-1}(\mathsf{a}_i) - \theta_i.
\label{eq:def_alpha_beta}
$
Consequently, for \eqref{eq: friction eq 1}:
\begin{align}
\label{eq: integration friction f0}
\sigma_{\text{DF}}^{(0)}(\mathsf{x}_i^{\prime})& :=  \frac{1}{2\pi} \int_{-\theta_i}^{2\pi-\theta_i} \sigma_{\text{DF}}\big(\Bar{\mathsf{v}}_i+\Tilde{\mathsf{v}}_i \cos(\omega \mathsf{t} +\theta_i)\big)d(\omega\mathsf{t}) \nonumber \\
& = \frac{1}{2\pi}\int_{-\theta_i}^{2\pi-\theta_i} \sigma_{\text{DF}}\big(\mathsf{a}_i+\cos(\omega \mathsf{t} +\theta_i)\big)d(\omega\mathsf{t}) \nonumber \\
& \stackrel{(a)}{=}  \frac{-\Delta}{2\pi} \Big( \int_{-\theta_i}^{\alpha_i}   + \int_{\beta_i}^{2\pi-\theta_i}   \Big) d(\omega \mathsf{t})
    + \frac{1}{2 \pi} \int_{\alpha_i}^{\beta_i} d(\omega \mathsf{t})
     \nonumber \\
    & = \frac{1}{\pi}\big(-\Delta\pi+(1+\Delta)\cos^{-1}(\mathsf{a}_i)\big),
\end{align}
where $(a)$ follows by using \eqref{eq:useful_a_region}.
Using similar arguments for \eqref{eq: friction eq 2}-\eqref{eq: friction eq 3}: 
\begin{align} \label{eq: appendix integrate f1}
    \sigma_{\text{DF}}^{(j)}(\mathsf{x}_i^{\prime})  :&= \frac{1}{\pi}\int_{-\theta_i}^{2\pi-\theta_i}\sigma_{\text{DF}}\big(\mathsf{a}_i+\cos(\omega\mathsf{t}+\theta_i)\big)\mathsf{b}_j(\omega\mathsf{t})d(\omega\mathsf{t}) \nonumber \\
    & = \frac{2}{\pi}(1+\Delta)\mathsf{b}_j(\pi-\theta_i)\sqrt{1-\mathsf{a}_i^2},
\end{align}
with
$
\mathsf{b}_j(\cdot) := 
   \begin{cases} 
   \cos(\cdot), \; & \text{ if } j = 1,\\
   \sin(\cdot), \; & \text{ if } j = 2.
   \end{cases}$

\subsection{Harmonic balance}
\label{subsec:harmonic_balance_appendix}
From \eqref{eq: strain equation} and \eqref{eq: velocity form}, we have $\mathsf{s}' = \mathsf{x}'_1- \mathsf{x}'_2$, i.e.,
$
    \omega \mathsf{A} \cos(\omega \mathsf{t}) = \Bar{\mathsf{v}}_1 - \Bar{\mathsf{v}}_2 + \Tilde{\mathsf{v}}_1\cos(\omega \mathsf{t} + \theta_1) - \Tilde{\mathsf{v}}_2\cos(\omega \mathsf{t}+ \theta_2). 
$
Using \textit{harmonic balance} in the different terms, yields
\begin{subequations}
    \begin{align}
         \Bar{\mathsf{v}}_1  & = \Bar{\mathsf{v}}_2, \label{eq: compatibility 1}\\
         \omega \mathsf{A} & = \Tilde{\mathsf{v}}_1\cos(\theta_1) - \Tilde{\mathsf{v}}_2\cos(\theta_2), \label{eq: compatibility 2}\\
         0 &= \Tilde{\mathsf{v}}_2\sin(\theta_2) -\Tilde{\mathsf{v}}_1 \sin(\theta_1). \label{eq: compatibility 3}
    \end{align}
\end{subequations}
\par\noindent Given \eqref{eq: compatibility 1}, for conciseness we write $\Bar{\mathsf{v}} = \Bar{\mathsf{v}}_i$.
Subtraction of \eqref{eq: dynamics 1} and \eqref{eq: dynamics 2} provides the strain dynamics
\begin{equation*}
    \mathsf{s}'' = 2\, \pi_\nd{f} \mathsf{f} + \pi_\sigma \big(\sigma_{\text{DF}}(\mathsf{x}_1')-\sigma_{\text{DF}}(\mathsf{x}_2')\big) 
     - \mathsf{s} - 2 \zeta \mathsf{s}'.
\end{equation*}
Using \eqref{eq: strain equation} together with \eqref{eq:describing_fcn_friction}-\eqref{eq:friction_coeffs}, \textit{harmonic balance }yields
\begin{subequations}
    \begin{align}
        0  & = \sigma_{\text{DF}}^{(0)}(\mathsf{x}_1') - \sigma_{\text{DF}}^{(0)}(\mathsf{x}_2') \label{eq: strain dynamics 1}\\
        0 & = 2\pi_\nd{f} \sin\phi+\pi_\sigma\big(\sigma_{\text{DF}}^{(1)}(\mathsf{x}_1')-\sigma_{\text{DF}}^{(1)}(\mathsf{x}_2')\big) - 2\zeta \mathsf{A} \omega \label{eq: strain dynamics 2}\\
        -\mathsf{A} \omega^2 & = 2\pi_\nd{f} \cos\phi +\pi_\sigma\big(\sigma_{\text{DF}}^{(2)}(\mathsf{x}_1')-\sigma_{\text{DF}}^{(2)}(\mathsf{x}_2')\big)- \mathsf{A} \label{eq: strain dynamics 3}
    \end{align}
\end{subequations}
\par\noindent Using \eqref{eq: friction eq 1} in  \eqref{eq: strain dynamics 1} and \eqref{eq: compatibility 3} sequentially, yields
\begin{equation}
    \mathsf{a}_1 = \mathsf{a}_2 \; \Rightarrow \; \Tilde{\mathsf{v}}_1 = \Tilde{\mathsf{v}}_2 \; \Rightarrow \; \sin\theta_1  = \sin\theta_2. \label{eq: v_tilde equality}
\end{equation}
Moving forward, we denote $\mathsf{a} = \mathsf{a}_i$ and $\Tilde{\mathsf{v}}  = \Tilde{\mathsf{v}}_i$. 
\eqref{eq: v_tilde equality} yields two possibilities: either $\theta_2 = \theta_1$ or $\theta_2 = \pi - \theta_1$. Since the two segments are not in phase during crawling, we must have
$
    \theta_2 = \pi - \theta_1.
$
Substitution 
in  \eqref{eq: compatibility 2} provides
\begin{equation}
    \tilde{\mathsf{v}} = \frac{\omega \mathsf{A} }{2\cos\theta_1}. \label{eq: V_bar A relation}
\end{equation}
Using \eqref{eq: friction eq 2}-\eqref{eq: friction eq 3} in \eqref{eq: strain dynamics 2}-\eqref{eq: strain dynamics 3} yields
\begin{subequations}
    \begin{align}
     \cos \phi & = \frac{\mathsf{A}}{2\pi_\nd{f}}  (1-\omega^2), \label{eq: phi A relation 2}\\
        \sin \phi & = \frac{1}{\pi_\nd{f}}\Big(\zeta \mathsf{A} \omega + 2 \pi_\sigma \frac{1+\Delta} {\pi}\sqrt{1-\mathsf{a}^2}\cos(\theta_1)\Big). \label{eq: phi A relation 1}
    \end{align}
        \label{eq:phi_and_A}
\end{subequations}
\par\noindent The dynamics  
$
    \mathsf{x}_1''+\mathsf{x}_2'' = \pi_\sigma(\sigma_{\text{DF}}(\mathsf{x}_1')+\sigma_{\text{DF}}(\mathsf{x}_2'))
$
provide the following identities through \textit{harmonic balance}
\begin{subequations}
    \begin{align}
        \mathsf{a} & = \cos\bigg(\frac{\Delta \pi}{1+\Delta}\bigg), \label{eq: COM dynamics 1}\\
         \sin\theta_1 & = 0.
        \label{eq: COM dynamics 2}
    \end{align}
\end{subequations}
\par\noindent Thus, there are two possible solutions for the phases of the nodal speeds: $\{\theta_1 = 0, \theta_2 = \pi\}$ and $\{\theta_1 = \pi, \theta_2 = 0\}$. Considering \eqref{eq: V_bar A relation} together with $\tilde{\mathsf{v}}>0$ and $\mathsf{A}>0$ $\Rightarrow \; \{\theta_1 = 0, \theta_2 = \pi\}$.
 With the values of $\theta_1$ and $\mathsf{a}$, \eqref{eq:phi_and_A} can be solved for $\phi$ and $\mathsf{A}$. \eqref{eq: V_bar A relation} provides the value of $\tilde{\mathsf{v}}$. Finally, $\Bar{\mathsf{v}}$ is obtained as $\Bar{\mathsf{v}} = \tilde{\mathsf{v}}\mathsf{a}$.
It is also noteworthy that the DC component of the frictional force is $\sigma_{\text{DF}}^{(0)}(\mathsf{x}_i') \equiv 0$ by \eqref{eq: COM dynamics 1}.

\end{document}